\documentstyle[ijmpa1,twoside]{article}

\input epsf.sty
\font\eightrm=cmr8
\renewcommand{\bar}[1]{\overline{#1}}

\newcommand{\etal}{{\em et al.}}

\begin{document}

\normalsize\textlineskip
\thispagestyle{empty}
\setcounter{page}{1}

\vspace*{0.88truein}

\fpage{1}
\centerline{\bf HIGH ENERGY PHOTON-PHOTON AND }
\vskip .1in
\centerline{{\bf ELECTRON-PHOTON COLLISIONS}
\footnote{Work supported by the Department of Energy under contract number
DE--AC03--76SF00515.}}
\vspace*{0.37truein}
\centerline{\footnotesize STANLEY J. BRODSKY}
\vspace*{0.015truein}
\centerline{\footnotesize\it Stanford Linear Accelerator Center}
\baselineskip=10pt
\centerline{\footnotesize\it Stanford University, Stanford, California 94309}
\vspace*{0.225truein}

\abstracts{The advent of a next linear $e^\pm e^-$ collider
and back-scatterd laser beams will allow the study
of a vast array of high energy processes of the Standard Model
through the fusion of real and virtual photons and other gauge
bosons. As examples, I discuss virtual photon scattering   $\gamma^* \gamma^*
\to X$ in the region dominated by  BFKL hard Pomeron  exchange   and report
the predicted cross sections at present and  future $e^\pm e^-$
colliders. I also discuss exclusive
$\gamma
\gamma$  reactions in QCD as a measure of hadron distribution amplitudes
and a new  method
for measuring the anomalous magnetic and quadrupole moments of the $W$ and
$Z$ gauge
bosons to high precision in polarized electron-photon collisions. }{}{}

\vspace*{1pt}\textlineskip	
\section{Introduction}
\vspace*{-0.5pt}
\noindent
The largest production rates at a high energy $e^\pm e^-$ collider arise
from photon-photon
fusion processes $e e \to e e \gamma \gamma \to e e X$ since the cross
sections increase
logarithmically with the available energy.\cite{bkt} The final state $X$ can
be hadrons, leptons, gauge bosons, and any other $C=+$ state coupling to the
electromagnetic current.  At very high energies the fusion of any pair of
gauge bosons
$\gamma, Z^0, W^\pm$  becomes accessible.
The final-state leptons in $e e
\to  e e X$ can be tagged to provide a source of virtual photons or $Z^0$
bosons of
tuneable energy and virtuality. In addition, by the use of laser beams
backscattering on a
polarized electron beam, one can obtain remarkably high luminosities for high
energy polarized real photon
$\gamma \gamma$ and $\gamma e$ collisions.\cite{Telnov}  Such collisions
will open up a large array of important tests of the Standard Model as well
as discovery
tools for new particles.

In this talk I will focus on
three important areas of physics involving high energy real and virtual
photon beams:
(a) exclusive two photon processes as a study of hadron structure in QCD,
(b) the study of
the total virtual photon-photon scattering cross section as  a definitive
probs of the
hard QCD (BFKL) pomeron; and (c) the study of of polarized-photon
polarized-electron
collisions in the process
$\gamma e^- \to W
\nu$ as a high precision test of the Standard model. The latter reaction is
particularly
well-matched to a high lumionosity polarized electron-electron and
backscattered laser
beam facility at the next linear collider.  High energy photon-photon
collisions also open up a huge range of novel QCD studies, such as
measurements of the
photon structure function, the search for $C = -1$ odderon exchange in
exclusive reactions
such as $\gamma \gamma \to \pi^0 \pi^0$  at $s \gg -t$,  and gluon jet
studies in
inclusive  reactions such as $\gamma \gamma \to g g$.

\vspace*{1pt}\textlineskip	
\section{Exclusive Photon-Photon Reactions}
\vspace*{-0.5pt}
\noindent
Exclusive $\gamma\gamma \rightarrow$ hadron pairs are among the most
fundamental processes in QCD, providing a detailed examination of
Compton scattering in the crossed channel.  In the high momentum
transfer domain $(s,t, \mbox{large},\theta_{cm}$ for $t/s$ fixed),
these processes can be computed from first principles in QCD, yielding
important information on the nature of the QCD coupling $\alpha_s$ and
the form of hadron distribution amplitudes.  Similarly, the transition
form factors $\gamma^*\gamma$, $\gamma^*\gamma \rightarrow \pi^0,
\eta^0,\eta^\prime,\eta_c \ldots$ provide rigorous tests of QCD and
definitive determinations of the meson distribution amplitudes
$\phi_H(x,Q)$.\cite{BrodskyLepage}

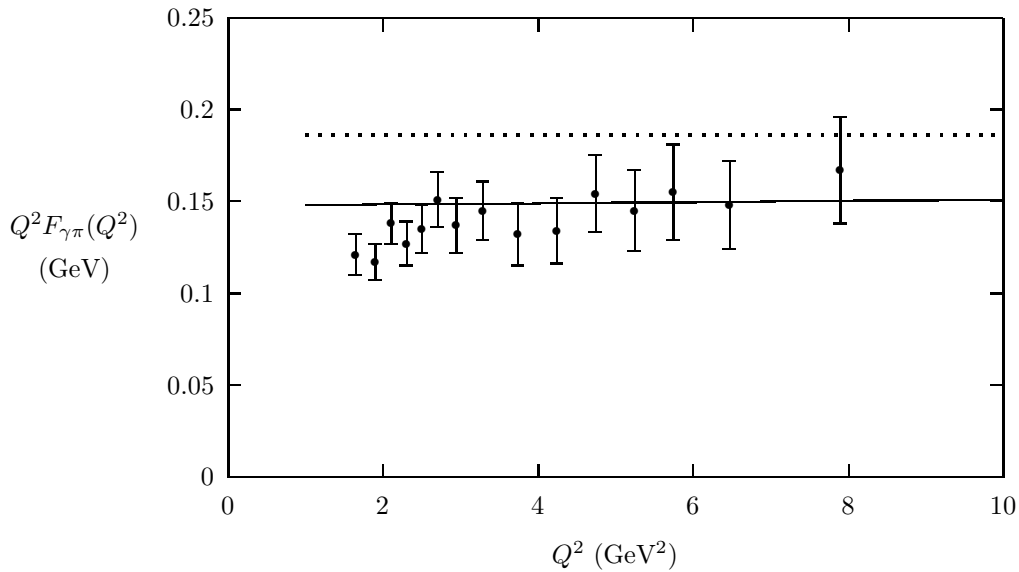
\begin{figure}[htb]
\begin{center}
\hbox to 4in{% GNUPLOT: LaTeX picture
\setlength{\unitlength}{0.240900pt}
\ifx\plotpoint\undefined\newsavebox{\plotpoint}\fi
\sbox{\plotpoint}{\rule[-0.200pt]{0.400pt}{0.400pt}}%
\begin{picture}(1500,900)(0,0)
\font\gnuplot=cmr10 at 10pt
\gnuplot
\sbox{\plotpoint}{\rule[-0.200pt]{0.400pt}{0.400pt}}%
\put(219.0,134.0){\rule[-0.200pt]{4.818pt}{0.400pt}}
\put(197,134){\makebox(0,0)[r]{0}}
\put(1416.0,134.0){\rule[-0.200pt]{4.818pt}{0.400pt}}
\put(219.0,278.0){\rule[-0.200pt]{4.818pt}{0.400pt}}
\put(197,278){\makebox(0,0)[r]{0.05}}
\put(1416.0,278.0){\rule[-0.200pt]{4.818pt}{0.400pt}}
\put(219.0,422.0){\rule[-0.200pt]{4.818pt}{0.400pt}}
\put(197,422){\makebox(0,0)[r]{0.1}}
\put(1416.0,422.0){\rule[-0.200pt]{4.818pt}{0.400pt}}
\put(219.0,567.0){\rule[-0.200pt]{4.818pt}{0.400pt}}
\put(197,567){\makebox(0,0)[r]{0.15}}
\put(1416.0,567.0){\rule[-0.200pt]{4.818pt}{0.400pt}}
\put(219.0,711.0){\rule[-0.200pt]{4.818pt}{0.400pt}}
\put(197,711){\makebox(0,0)[r]{0.2}}
\put(1416.0,711.0){\rule[-0.200pt]{4.818pt}{0.400pt}}
\put(219.0,855.0){\rule[-0.200pt]{4.818pt}{0.400pt}}
\put(197,855){\makebox(0,0)[r]{0.25}}
\put(1416.0,855.0){\rule[-0.200pt]{4.818pt}{0.400pt}}
\put(219.0,134.0){\rule[-0.200pt]{0.400pt}{4.818pt}}
\put(219,89){\makebox(0,0){0}}
\put(219.0,835.0){\rule[-0.200pt]{0.400pt}{4.818pt}}
\put(462.0,134.0){\rule[-0.200pt]{0.400pt}{4.818pt}}
\put(462,89){\makebox(0,0){2}}
\put(462.0,835.0){\rule[-0.200pt]{0.400pt}{4.818pt}}
\put(706.0,134.0){\rule[-0.200pt]{0.400pt}{4.818pt}}
\put(706,89){\makebox(0,0){4}}
\put(706.0,835.0){\rule[-0.200pt]{0.400pt}{4.818pt}}
\put(949.0,134.0){\rule[-0.200pt]{0.400pt}{4.818pt}}
\put(949,89){\makebox(0,0){6}}
\put(949.0,835.0){\rule[-0.200pt]{0.400pt}{4.818pt}}
\put(1193.0,134.0){\rule[-0.200pt]{0.400pt}{4.818pt}}
\put(1193,89){\makebox(0,0){8}}
\put(1193.0,835.0){\rule[-0.200pt]{0.400pt}{4.818pt}}
\put(1436.0,134.0){\rule[-0.200pt]{0.400pt}{4.818pt}}
\put(1436,89){\makebox(0,0){10}}
\put(1436.0,835.0){\rule[-0.200pt]{0.400pt}{4.818pt}}
\put(219.0,134.0){\rule[-0.200pt]{293.175pt}{0.400pt}}
\put(1436.0,134.0){\rule[-0.200pt]{0.400pt}{173.689pt}}
\put(219.0,855.0){\rule[-0.200pt]{293.175pt}{0.400pt}}
\put(-22,494){\makebox(0,0){\shortstack{$Q^2 F_{\gamma\pi} (Q^2)$\\ \\ (GeV)}}}
\put(827,10){\makebox(0,0){$Q^2$ (GeV$^2$)}}
\put(219.0,134.0){\rule[-0.200pt]{0.400pt}{173.689pt}}
\put(419,483){\circle*{12}}
\put(450,471){\circle*{12}}
\put(475,532){\circle*{12}}
\put(499,500){\circle*{12}}
\put(523,523){\circle*{12}}
\put(548,569){\circle*{12}}
\put(577,529){\circle*{12}}
\put(619,552){\circle*{12}}
\put(674,515){\circle*{12}}
\put(735,520){\circle*{12}}
\put(796,578){\circle*{12}}
\put(857,552){\circle*{12}}
\put(918,581){\circle*{12}}
\put(1006,561){\circle*{12}}
\put(1180,616){\circle*{12}}
\put(419.0,451.0){\rule[-0.200pt]{0.400pt}{15.418pt}}
\put(409.0,451.0){\rule[-0.200pt]{4.818pt}{0.400pt}}
\put(409.0,515.0){\rule[-0.200pt]{4.818pt}{0.400pt}}
\put(450.0,443.0){\rule[-0.200pt]{0.400pt}{13.731pt}}
\put(440.0,443.0){\rule[-0.200pt]{4.818pt}{0.400pt}}
\put(440.0,500.0){\rule[-0.200pt]{4.818pt}{0.400pt}}
\put(475.0,500.0){\rule[-0.200pt]{0.400pt}{15.418pt}}
\put(465.0,500.0){\rule[-0.200pt]{4.818pt}{0.400pt}}
\put(465.0,564.0){\rule[-0.200pt]{4.818pt}{0.400pt}}
\put(499.0,466.0){\rule[-0.200pt]{0.400pt}{16.622pt}}
\put(489.0,466.0){\rule[-0.200pt]{4.818pt}{0.400pt}}
\put(489.0,535.0){\rule[-0.200pt]{4.818pt}{0.400pt}}
\put(523.0,486.0){\rule[-0.200pt]{0.400pt}{18.067pt}}
\put(513.0,486.0){\rule[-0.200pt]{4.818pt}{0.400pt}}
\put(513.0,561.0){\rule[-0.200pt]{4.818pt}{0.400pt}}
\put(548.0,526.0){\rule[-0.200pt]{0.400pt}{20.958pt}}
\put(538.0,526.0){\rule[-0.200pt]{4.818pt}{0.400pt}}
\put(538.0,613.0){\rule[-0.200pt]{4.818pt}{0.400pt}}
\put(577.0,486.0){\rule[-0.200pt]{0.400pt}{20.717pt}}
\put(567.0,486.0){\rule[-0.200pt]{4.818pt}{0.400pt}}
\put(567.0,572.0){\rule[-0.200pt]{4.818pt}{0.400pt}}
\put(619.0,506.0){\rule[-0.200pt]{0.400pt}{22.163pt}}
\put(609.0,506.0){\rule[-0.200pt]{4.818pt}{0.400pt}}
\put(609.0,598.0){\rule[-0.200pt]{4.818pt}{0.400pt}}
\put(674.0,466.0){\rule[-0.200pt]{0.400pt}{23.608pt}}
\put(664.0,466.0){\rule[-0.200pt]{4.818pt}{0.400pt}}
\put(664.0,564.0){\rule[-0.200pt]{4.818pt}{0.400pt}}
\put(735.0,469.0){\rule[-0.200pt]{0.400pt}{24.813pt}}
\put(725.0,469.0){\rule[-0.200pt]{4.818pt}{0.400pt}}
\put(725.0,572.0){\rule[-0.200pt]{4.818pt}{0.400pt}}
\put(796.0,518.0){\rule[-0.200pt]{0.400pt}{29.149pt}}
\put(786.0,518.0){\rule[-0.200pt]{4.818pt}{0.400pt}}
\put(786.0,639.0){\rule[-0.200pt]{4.818pt}{0.400pt}}
\put(857.0,489.0){\rule[-0.200pt]{0.400pt}{30.594pt}}
\put(847.0,489.0){\rule[-0.200pt]{4.818pt}{0.400pt}}
\put(847.0,616.0){\rule[-0.200pt]{4.818pt}{0.400pt}}
\put(918.0,506.0){\rule[-0.200pt]{0.400pt}{36.135pt}}
\put(908.0,506.0){\rule[-0.200pt]{4.818pt}{0.400pt}}
\put(908.0,656.0){\rule[-0.200pt]{4.818pt}{0.400pt}}
\put(1006.0,492.0){\rule[-0.200pt]{0.400pt}{33.244pt}}
\put(996.0,492.0){\rule[-0.200pt]{4.818pt}{0.400pt}}
\put(996.0,630.0){\rule[-0.200pt]{4.818pt}{0.400pt}}
\put(1180.0,532.0){\rule[-0.200pt]{0.400pt}{40.230pt}}
\put(1170.0,532.0){\rule[-0.200pt]{4.818pt}{0.400pt}}
\put(1170.0,699.0){\rule[-0.200pt]{4.818pt}{0.400pt}}
\sbox{\plotpoint}{\rule[-0.500pt]{1.000pt}{1.000pt}}%
\put(341,670){\usebox{\plotpoint}}
\multiput(341,670)(20.756,0.000){52}{\usebox{\plotpoint}}
\put(1418,670){\usebox{\plotpoint}}
\sbox{\plotpoint}{\rule[-0.200pt]{0.400pt}{0.400pt}}%
\put(341,561){\usebox{\plotpoint}}
\put(430,560.67){\rule{5.300pt}{0.400pt}}
\multiput(430.00,560.17)(11.000,1.000){2}{\rule{2.650pt}{0.400pt}}
\put(341.0,561.0){\rule[-0.200pt]{21.440pt}{0.400pt}}
\put(587,561.67){\rule{5.300pt}{0.400pt}}
\multiput(587.00,561.17)(11.000,1.000){2}{\rule{2.650pt}{0.400pt}}
\put(452.0,562.0){\rule[-0.200pt]{32.521pt}{0.400pt}}
\put(698,562.67){\rule{5.541pt}{0.400pt}}
\multiput(698.00,562.17)(11.500,1.000){2}{\rule{2.770pt}{0.400pt}}
\put(609.0,563.0){\rule[-0.200pt]{21.440pt}{0.400pt}}
\put(810,563.67){\rule{5.300pt}{0.400pt}}
\multiput(810.00,563.17)(11.000,1.000){2}{\rule{2.650pt}{0.400pt}}
\put(721.0,564.0){\rule[-0.200pt]{21.440pt}{0.400pt}}
\put(944,564.67){\rule{5.541pt}{0.400pt}}
\multiput(944.00,564.17)(11.500,1.000){2}{\rule{2.770pt}{0.400pt}}
\put(832.0,565.0){\rule[-0.200pt]{26.981pt}{0.400pt}}
\put(1056,565.67){\rule{5.300pt}{0.400pt}}
\multiput(1056.00,565.17)(11.000,1.000){2}{\rule{2.650pt}{0.400pt}}
\put(967.0,566.0){\rule[-0.200pt]{21.440pt}{0.400pt}}
\put(1190,566.67){\rule{5.300pt}{0.400pt}}
\multiput(1190.00,566.17)(11.000,1.000){2}{\rule{2.650pt}{0.400pt}}
\put(1078.0,567.0){\rule[-0.200pt]{26.981pt}{0.400pt}}
\put(1324,567.67){\rule{5.541pt}{0.400pt}}
\multiput(1324.00,567.17)(11.500,1.000){2}{\rule{2.770pt}{0.400pt}}
\put(1212.0,568.0){\rule[-0.200pt]{26.981pt}{0.400pt}}
\put(1347.0,569.0){\rule[-0.200pt]{21.440pt}{0.400pt}}
\end{picture}}
\end{center}
\caption[*]{The $\gamma\rightarrow\pi^0$ transition form factor.  The
solid line is the full prediction including the QCD correction; the dotted line
is the LO prediction $Q^2F_{\gamma\pi}(Q^2) = 2f_\pi$.}
\label{fgammapi}
\end{figure}

The simplest hadronic exclusive reaction is the  $\gamma e \to
e^\prime \pi^0$ process which measures the $\gamma \to \pi^0$ transition
form factor. The present data from CLEO shown in
figure (1) shows remarkable consistency with the next-to-leading-order
(NLO) leading-twist
scaling QCD predictions \cite {BrodskyLepage,Braaten} for photon
virtualities up to
$Q^2$ = 8
$GeV^2$.  See Fig. (\ref{fgammapi}). The consistency of the CLEO data
\cite{CLEO}  with
the predicted normalization and the apparent absence of violations of
leading-twist scaling
requires that the basic wavefunction which describes the momentum
distribution in the valence
$q
\bar q$ Fock state of the pion, the pion distribution amplitude,  has the form
$\phi(x,Q) =
\sqrt 3 f_\pi x (1-x)$, which is the asymptotic leading anomalous dimension solution
to the QCD evolution equation for the meson light-cone wavefunction.  It also
assumes that the running coupling which appears in the NLO corrections is
slowly varying at small momentum transfer.\cite {BJPR}

Since the cross section of
the single meson exclusive reaction falls off as only one power of $Q^2$
compared to
scale-invariant reactions, studies of the normalization and
scaling of other
$\gamma e \to e^\prime M^0$ reactions should be practical for
general
$C=+$ neutral mesons including the $\eta_c$ and $ \eta_b$ at high photon
virtuality $Q^2$
in high energy
$\gamma e$ collisions. Such measurements can provide fundamental
information on the nature
of the hadron wavefunctions and their anomalous dimensions. Further
discussion of the
theory of exclusive single hadron and hadron pair production in
$\gamma\gamma$ and $e \gamma$ collisions can be found in a recent paper by Ji,
Robertson, Pang and myself.\cite {BJPR}

\vspace*{1pt}\textlineskip	
\section{Probing the Hard QCD Pomeron in Virtual Photon Collisions}
\vspace*{-0.5pt}
\noindent
The BFKL equation describes scattering processes in QCD in the limit of
large energies  and fixed (sufficiently large) momentum transfers. The
study discussed in this section analyzes the prospects for using
virtual photon-photon collisions at LEPII and a next linear collider as a probe
of QCD dynamics in this region.\cite {bhshep,bartels} The quantity we focus on
is the total cross section for scattering two photons sufficiently far off shell
at large center-of-mass energies, $
\gamma^* (Q_A^2) + \gamma^* (Q_B^2) \to {\mbox {hadrons}} $, $ s \gg
Q_A^2 , Q_B^2 \gg \Lambda^2_{ QCD}$. This process can be observed at
high-energy and high-luminosity $e^\pm e^-$ colliders as well as $\mu^\pm \mu^-$
colliders, where the photons are produced from the lepton beams by
bremsstrahlung. The $\gamma^* \gamma^*$ cross section can be measured in
collisions in which both the outgoing leptons are tagged.

The basic motivation  for this study is that compared to tests of BFKL
dynamics in deeply inelastic lepton-hadron scattering (see, for
instance, the review by Abramowicz \cite{abra}) the off-shell photon cross
section presents an essential theoretical advantage because it
does not involve a non-perturbative target. The photons act as color
dipoles with small transverse size, so that the QCD interactions can be
treated in a fully perturbative framework.

\begin{figure}[htb]
\begin{center}
\leavevmode
 {\epsfxsize 3in\epsfbox{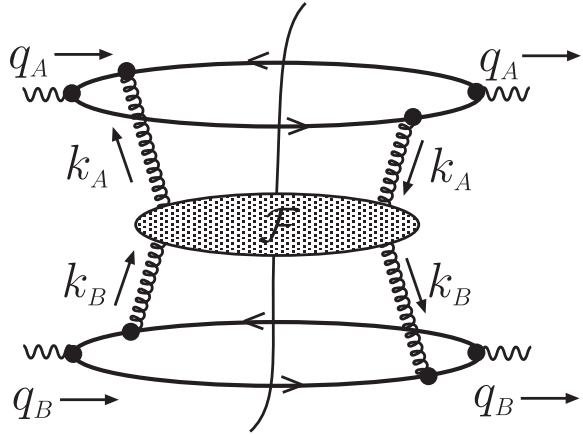}}
\end{center}
\caption[*]{The virtual photon cross section
in the high energy limit.
\label{fac}}
\end{figure}

The structure of $\gamma^* \gamma^*$ high-energy scattering is shown
schematically in Fig.~(\ref{fac}). We work in a frame in which the photons $q_A,
\, q_B$ have zero transverse momenta and are boosted along the positive
and negative light-cone directions. In the leading logarithm
approximation, the  process can be described as the interaction of two
$q {\bar q}$ pairs scattering off each other via multiple gluon
exchange. The $q {\bar q}$ pairs are in a color-singlet state and
interact through their color dipole moments. The gluonic function
${\cal F}$ is obtained from the solution to the BFKL
equation.\cite{bfkl}
\vspace{.5cm}

The analysis of the transverse-distance scales involved in the
scattering illustrates a few distinctive features of this process. The
mean transverse size of each $q {\bar q}$ dipole is given, in the first
approximation, by the reciprocal of the corresponding photon virtuality:
\begin{equation}
\label{meanrt}
<R_{\perp \, A}> \, \sim 1 / Q_A \hspace*{0.4 cm} , \hspace*{0.6 cm}
<R_{\perp \, B}> \, \sim 1 / Q_B \hspace*{0.6 cm} .
\end{equation}
However, fluctuations can bring in much larger transverse sizes.
Large-size fluctuations occur as a result of the configurations in
which one quark of the pair carries small transverse momentum and a
small fraction of  the photon longitudinal momentum (the so-called
aligned-jet configurations~\cite{aligned}) . For example, for the
momentum $p_A$ of the quark created by photon $A$: \begin{equation}
\label{alicfg} {\bf p}_{\perp \, A} \ll  Q_A \hspace*{0.4 cm} ,
\hspace*{0.6 cm} z_A \equiv p_A^{+}/q_A^{+} \ll 1  \hspace*{0.6 cm} .
\end{equation} The actual size up to which the $q {\bar q}$ pair can
fluctuate is controlled by the scale of the system that it scatters
off. Therefore, in $\gamma^* \gamma^*$ scattering the fluctuations  in
the transverse size of each pair are suppressed by the off-shellness of
the  photon creating the other pair. If {\em both} photons are
sufficiently far off shell, the transverse separation in each $q {\bar
q}$ dipole stays small.\cite{bhshep}  This can be contrasted with the
case of deeply inelastic $e \, p$ scattering (or $e \, \gamma$, where
$\gamma$ is a (quasi-)real photon). In this case, the  $q {\bar q}$
pair produced by the virtual photon can fluctuate up to sizes of the
order of a hadronic scale, that is, $1 / \Lambda_{QCD}$. This results
in the deeply inelastic  cross section being determined by an interplay
of short  and long distances.

\vspace{.5cm}
\begin{figure}[htb]
\begin{center}
\leavevmode
{\epsfxsize 4in \epsfbox{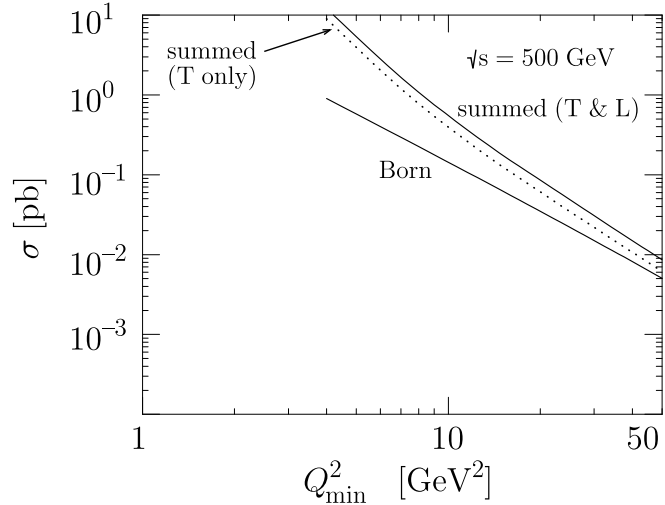}}
\end{center}
\caption[*]{ The $Q_{\rm min}^2$ dependence of the $e^+ e^-$
integrated rate for $\protect\sqrt s = 500\ {\rm GeV}$. The choice of
the cuts and of the scales in the leading logarithm result is as in
Brodsky \etal\cite{bhshep} The dot-dashed and solid lines correspond
to the result of using, respectively, the Born and the BFKL-summed
expressions for the photon-photon cross section. The dotted curve shows
the contribution to the summed result coming from transversely
polarized photons.
\label{f500}}
\end{figure}

\vspace{.5cm}
\begin{figure}[htbp]
\begin{center}
\leavevmode
 {\epsfxsize 4in \epsfbox{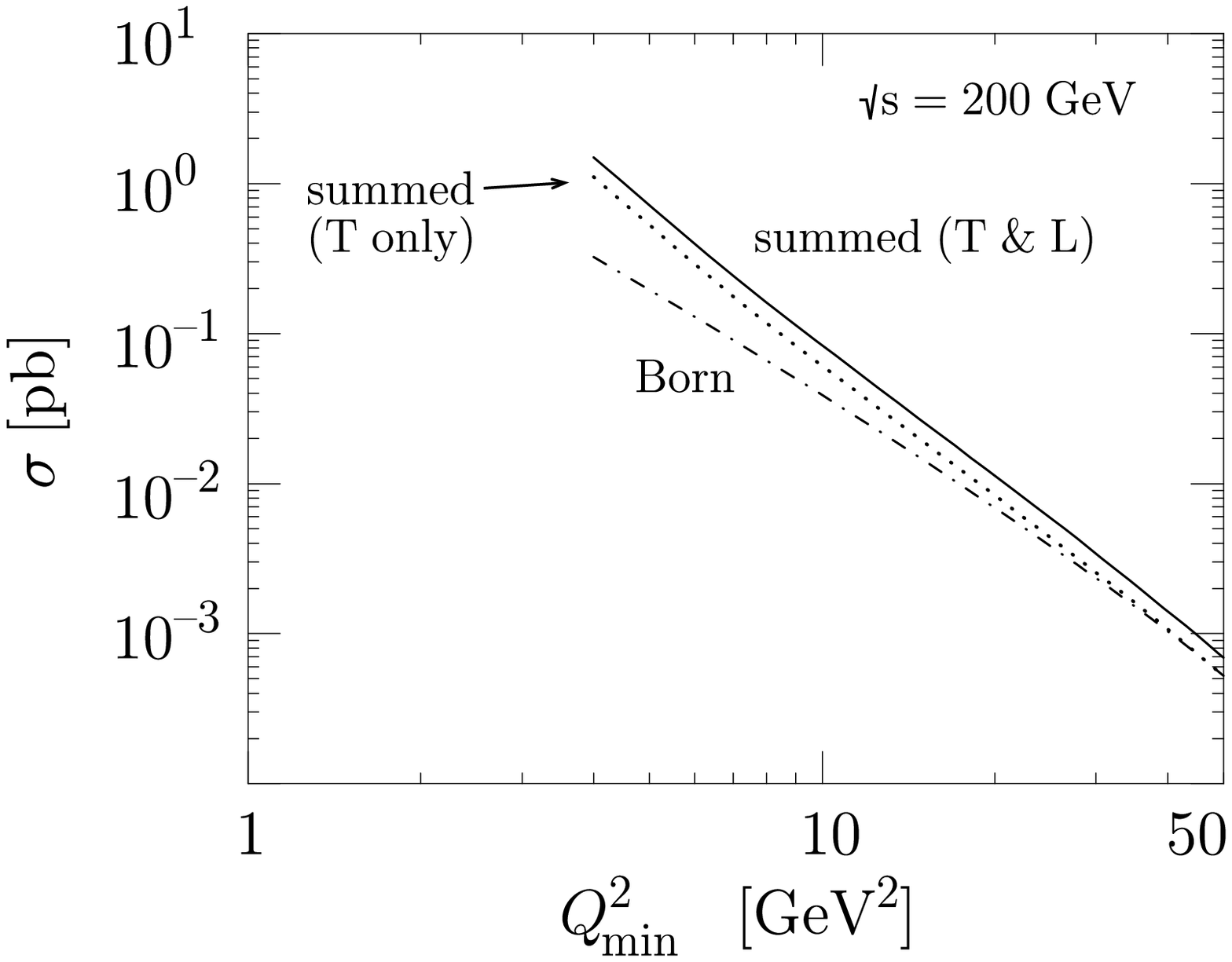}}
\end{center}
\caption[*]{Same as in
Fig.~\protect\ref{f500} for $\protect\sqrt s = 200\ {\rm GeV}$.
\label{f200}}
\end{figure}

In principle, the  $q {\bar q}$ dipoles in the $\gamma^* \gamma^*$
process could  still fluctuate to  bigger sizes in correspondence of
configurations in which the jet alignment occurs twice, once for each
photon. However, such configurations cost an extra overall power of $1
/ Q^2$ in the cross section (terms proportional to $1 / (Q_A^2 \, Q_B^2
) $ rather than $1 / (Q_A \, Q_B ) $).\cite{bjfutu} Therefore, they
only contribute at the level of sub-leading power corrections  to
$\sigma (\gamma^* \gamma^*)$.

Even though the $q {\bar q}$ dipoles have small transverse size,
sensitivity to large transverse distances may be brought in through the
BFKL function   ${\cal F}$. This    indeed is expected to  occur when
the energy $s$ becomes very large. As $s$ increases, the typical impact
parameters dominating the cross section for BFKL exchange grow to be
much larger than the size of the colliding objects.\cite{mdiff} One
can interpret this   as providing an upper bound on the range of values
of  $\left( \alpha_s (Q^2) \, \ln (s / Q^2) \right)$ in which the
simple BFKL approach to virtual photon  scattering is expected to give
reliable predictions.\cite{bhshep}

The calculation of $\sigma (\gamma^* \gamma^*)$ and the form of the
result are discussed in detail in recent papers.\cite {bhshep,bartels}. We
recall
here the main features:

i) for large virtualities,  $\sigma (\gamma^* \gamma^*)$ scales like
$1/Q^2$, where $Q^2 \sim {\mbox {max}} \{ Q_A^2 , Q_B^2\} $. This is
characteristic of the perturbative QCD prediction. Models based on
Regge factorization (which work well in the soft-interaction regime
dominating $\gamma \, \gamma$ scattering near the mass shell) would
predict a higher power in $1/Q$.

ii) $\sigma (\gamma^* \gamma^*)$ is affected by  logarithmic
corrections in the energy $s$ to all orders in $\alpha_s$. As a result
of the BFKL summation of these contributions, the cross section rises
like   a power in $s$, $\sigma \propto s^\lambda$. The Born
approximation to this result (that is, the ${\cal O} (\alpha_s^2) $
contribution, corresponding to  single gluon exchange in the graph of
Fig.~2) gives a constant cross section, $\sigma_{Born} \propto s^0$.
This behavior in $s$ can be compared with lower-order calculations
which do not include the  corrections associated to (single or
multiple) gluon exchange. Such calculations would give cross sections
that fall off like $1 / s$ at large $s$.

These features are reflected at the level of the  $e^\pm e^-$ scattering
process. The $e^\pm e^-$ cross section is obtained by folding  $ \sigma
(\gamma^{*} \, \gamma^{*}) $ with the flux of photons from each lepton.
In Figs.~\ref{f500} and \ref{f200}, we integrate  this cross section
with a lower cut on the photon virtualities (in order that the coupling
$\alpha_s$ be small, and that the process be dominated by the
perturbative contribution) and a lower cut on the photon-photon c.m.s.
energy (in order that the high energy approximation be valid). We plot
the result as a function of  the lower bound $Q_{\rm min}^2$,
illustrating the expected dependence of the photon-photon cross section
on the photon virtualities. Figure~\ref{f500} is for the energy of a
future $e^\pm e^-$ collider.  Figure~\ref{f200} refers to the LEP collider
operating at $\sqrt s = 200\ {\rm GeV}$. Details on our choice of cuts
may be found in my paper with Hautmann and Soper, \etal\cite{bhshep}

\begin{figure}[htbp]
\begin{center}
\leavevmode
{\epsfxsize 4in \epsfbox{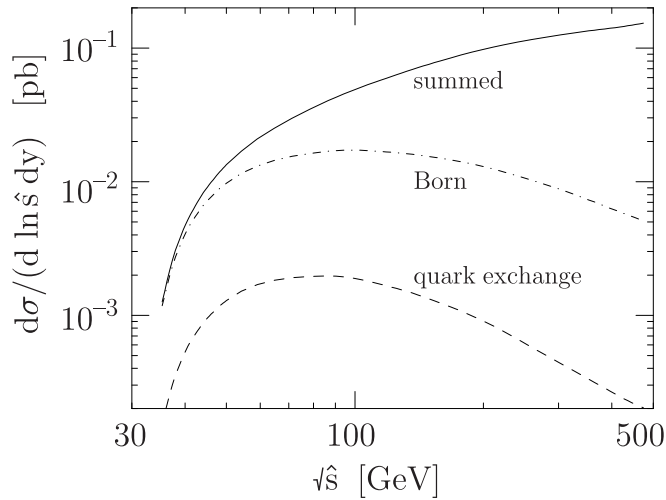}}
\end{center}
\caption[*]{The cross section $d\sigma / (d \ln\hat s\, dy)$ at $y = 0$
for $\protect\sqrt s = 500\ {\rm GeV}$. We take $Q_{\rm min}^2 = 10\
{\rm GeV}^2$. The solid curve is the summed BFKL result. The dot-dashed
curve is the Born result. The dashed curve shows the (purely
electromagnetic) contribution arising from the scattering of
(transversely polarized) photons via quark exchange.
\label{hats500}}
\end{figure}

From Figs.~\ref{f500} and \ref{f200}, for a value of the cut $Q_{\rm
min} = 2 \ {\mbox {GeV}}$ we find $ \sigma \simeq 1.5 \, {\mbox {pb}} $
at LEP200 energies, and $ \sigma \simeq 12 \, {\mbox {pb}} $ at the
energy of a future collider. These cross sections would give rise to
about $750$ events at LEP200 for a value of the luminosity $L = 500 \,
{\mbox {pb}}^{-1}$, and about $6 \times 10^5$ events at $\sqrt s = 500\
{\rm GeV}$ for $L = 50 \, {\mbox {fb}}^{-1}$. The above value of
$Q_{\rm min}$ would imply detecting leptons scattered through angles
down to about $20 \ {\mbox {mrad}}$ at LEP200, and about $8\ {\mbox
{mrad}}$ at a future $500 \, {\mbox {GeV}}$ collider.  If instead we
take, for instance, $Q_{\rm min} = 6\  {\rm GeV}$, the minimum angle at
a $500 \, {\mbox {GeV}}$ collider is $24\ {\mbox {mrad}}$. Then the
cross section is about $2 \times 10^{-2}\ {\rm pb}$, corresponding to
about $10^3$ events.
The dependence on the photon-photon c.m.\ energy $\sqrt {\hat s}$ can
be best studied by fixing $Q_{\rm min}$ and  looking at the cross
section $d\sigma /( d \ln\hat s\, dy)$ (here $y$ is the photon-photon
rapidity). In Figure~\ref{hats500} we plot this cross section at $y = 0$.
While at the lowest end of the range in $\sqrt{\hat s}$ the curves are
strongly dependent on the choice of the cuts, for increasing
$\sqrt{\hat s}$ the plotted distribution is rather directly related to
the behavior of $\sigma (\gamma^* \gamma^*)$ discussed earlier. In
particular, as $\sqrt{\hat s}$ increases to about $100 \ {\rm GeV}$ we
see the Born result flatten out and the summed BFKL result rise, while
the contribution from quark exchange is comparatively  suppressed.  The
damping  towards the higher end of the range in $\sqrt{\hat s}$ affects
all curves and is due to the influence of the photon flux factors.

\vspace{.5cm}

Figure~\ref{hats500} is computed for $\sqrt{s} = 500\ {\rm GeV}$. The
corresponding curves at LEP200 energies are qualitatively similar. The
main difference is that at $\sqrt{s} = 200\ {\rm GeV}$ there is less
available range for $\sqrt{\hat s}$.

We see from the results presented above that at a future $e^\pm e^-$
collider it should be possible to probe the effects of pomeron exchange
in a range of $Q^2$ where summed perturbation theory applies. One
should be able to investigate this region in detail by varying $Q_A$,
$Q_B$ and ${\hat s}$ independently. At LEP200 such studies appear to be
more problematic mainly because of limitations in luminosity. Even with
a modest luminosity, however, one can access the region of relatively
low $Q^2$ if one can get down to small enough angles. This would allow
one to examine experimentally the transition between soft and hard
scattering.

\vspace*{1pt}\textlineskip	
\section{Precision
limits on Anomalous Couplings of the $W$ and $Z$\cite{BRS}}
\vspace*{-0.5pt}
\noindent
The Dirac value $g = 2$ for the magnetic moment $\mu = {g e S /2M }$ of a
particle of charge $e$, mass $M$, and spin $S$, plays a special role in
quantum field theory.  As shown by Weinberg \cite{wein}  and Ferrara
{\it et. al} \cite{FPT}, the canonical value $g=2$ gives an effective
Lagrangian which has maximally convergent high energy behavior for fields
of any spin.  In the case of the Standard Model, the anomalous magnetic
moments $\mu_a = (g-2)e S / 2 M $ and anomalous quadrupole moments $Q_a =
Q + {e /M^2} $ of the fundamental fields vanish at tree level, ensuring a
quantum field theory which is perturbatively renormalizable.  However, as
one can use the DHG sum
rule \cite{drell}  to show that the magnetic and quadrupole moments of
spin-${1\over2}$ or spin-$1$ bound states approach the canonical values $\mu=
e S/ M $ and $Q = - e/M^2$ in the zero radius limit $M R \to
0$ \cite{BD,BSCHLUMPF,brod92}, independent of the internal dynamics.
Deviations from the predicted values will thus reflect new physics and
interactions such as virtual corrections from supersymmetry or an
underlying composite structure.

The canonical values $g=2$ and $Q = -e/M^2$ lead to a number of important
phenomenological consequences: (1) The magnetic moment of a particle with
$g=2$ processes with the same frequency as the Larmor frequency in a
constant magnetic field.  This synchronicity is a consequence of the fact
that the electromagnetic spin currents can be formally generated by an
infinitesimal Lorentz transformation. \cite{BMT,BBK} (2) The forward
helicity-flip Compton amplitude for a target with $g=2$ vanishes at zero
energy. \cite{Low} (3) The Born amplitude for a photon radiated in the
scattering of any number of incoming and outgoing particles with charge
$e_i$ and four-momentum $p_i^\mu$ vanishes at the kinematic angle where
all the ratios ${e_i/p_i\cdot k}$ are simultaneously equal.~\cite{BBK}
For example, the Born cross section ${d \sigma/\cos \theta_{cm}} (u
\overline
d \to W^+ \gamma)$ vanishes identically at an angle determined from the
ratio of charges: $\cos \theta_{cm} = {e_d/e_{W^+} =
-1/3}$.~\cite{MSS}  Such ``radiative amplitude zeroes" or ``null
zones" occur at lowest order in the Standard Model because the
electromagnetic spin currents of the quarks and the vector gauge bosons
are all canonical.

The vanishing of the forward helicity-flip Compton amplitude at zero
energy for the canonical couplings, together with the optical theorem and
dispersion theory, leads to a superconvergent sum rule; {\it i.e.}, a zero value
for the DHG sum rule.  This remarkable observation was first made for
quantum electrodynamics and the electroweak theory by Altarelli, Cabibbo
and Maiani. \cite{ACM}  More generally, one can use a quantum
loop \cite{BS}  expansion to show that the logarithmic integral of
the spin-dependent part of the photoabsorption cross section
\begin{equation}
\int^\infty_{\nu_{th}}{d\nu\over \nu} \Delta \sigma_{\rm Born}(\nu) = 0
\label{eqaa}
\end{equation}
for any $2 \to 2$ Standard Model process $\gamma a \to b c$
in the classical, tree graph approximation.  The particles $a, b, c$ and
$d$ can be leptons, photons, gluons, quarks, elementary Higgs particles,
supersymmetric particles, etc.  We also can extend the sum rule to
certain virtual photon processes.  Here $\nu = {p \cdot q}/M$ is the
laboratory energy and $\Delta \sigma(\nu) = \sigma_P(\nu)- \sigma_A(\nu)$
is the difference between the photoabsorption cross section for parallel
and antiparallel photon and target helicities.  The sum rule receives
nonzero contributions in higher order perturbation theory in the Standard
Model from both quantum loop corrections and higher particle number final
states.  Similar arguments also imply that the DHG integral vanishes for
virtual photoabsorption processes such as $\ell \gamma \to \ell Q
\overline Q$
and $\ell g \to \ell Q \overline Q,$ the lowest order sea-quark contribution
to polarized deep inelastic photon and hadron structure functions.  Note
that the integral extends to $\nu = \nu_{th},$ which is generally beyond
the usual leading twist domain.

\vspace{.5cm}
\begin{figure}[htbp]
\begin{center}
\leavevmode
{\epsfxsize=3.truein \epsfbox{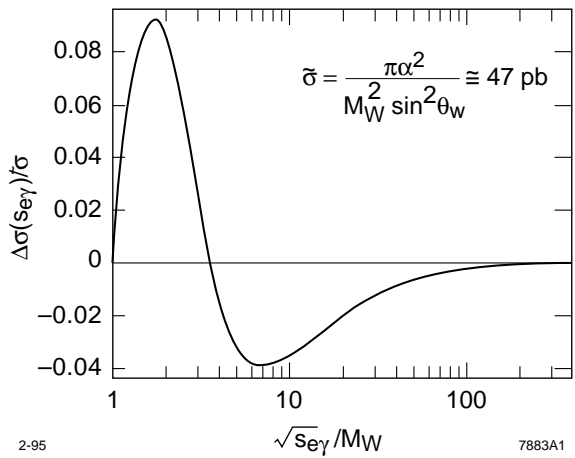}}
\end{center}
\caption[*]{The Born cross section difference $\Delta \sigma$ for the
Standard Model process $\gamma e \to W \nu$ for parallel minus
antiparallel electron/photon helicities as a function of $\log \sqrt
s_{e \gamma}/M_W$ The logarithmic integral of $\Delta \sigma$
vanishes in the classical limit.}
\label{figB}
\end{figure}

Schmidt, Rizzo and I \cite {BRS} have shown that one  can use Eq.
(\ref{eqaa}) as
a new way to test the canonical couplings of the Standard Model and to isolate
the higher order radiative corrections. The sum rule also provides a non-trivial
consistency check on calculations of the polarized cross sections.
Probably the most interesting application and test of the Standard
Model is to the reactions $\gamma \gamma \to q \overline q$, $\gamma e
\to W \nu$ and $\gamma e \to Z e$ which can be studied in high
energy polarized electron-positron colliders with back-scattered
laser beams.  In contrast to the timelike process $e^+ e^- \to W^+
W^-$, the $\gamma \gamma$ and $\gamma e$ reactions are sensitive to
the anomalous moments of the gauge bosons at $q^2 = 0.$ The
cancellation of the positive and negative contributions \cite{ginz}
of $\Delta \sigma(\gamma e \to W \nu)$ to the DHG integral is
evident in Fig.~\ref{figB}.

We can also exploit
the fact that the vanishing of the logarithmic integral of
$\Delta \sigma$ in the Born approximation also implies that there must be
a center-of-mass energy, $\sqrt s_0$, where the polarization asymmetry
$A=\Delta \sigma/ \sigma$ possesses a zero, {\it i.e.}, where $\Delta
\sigma({\gamma e \to W \nu })$ reverses sign. \cite{BRS}
We find strong sensitivity of the position of this zero or
``crossing point'' (which occurs at $\sqrt s_{\gamma e} =
3.1583 \ldots  M_W
\simeq 254$ GeV in the SM) to modifications of the SM trilinear $\gamma W
W$ coupling.  Given reasonable assumptions for the luminosity and energy
range for the Next Linear Collider(NLC), the zero point, $\sqrt s_0$, of
the polarization asymmetry may be determined with sufficient precision to
constrain the anomalous couplings of the $W$ to better than the 1\% level
at $95\%$ CL.  Since the zero occurs at rather modest energies where the
unpolarized cross section is near its maximum, an electron-positron
collider with ${\sqrt s}=320-400$ GeV is sufficient, whereas other
techniques aimed at probing the anomalous couplings through the $\gamma e
\to W \nu$ process require significantly larger energies.  In addition to
the fact that only a limited range of energy is required, the
polarization asymmetry measurements have the advantage that many
of the systematic errors cancel in taking cross section ratios.  This
technique can clearly be generalized to other higher order tree-graph
processes in the Standard Model and supersymmetric gauge theory.  The
position of the zero in the photoabsorption asymmetry thus provides an
additional weapon in the arsenal used to probe anomalous trilinear gauge
couplings.

\section{Acknowledgments}

\medskip
This work is supported in part by the United States Department of
Energy grants DE-AC03-76SF00515 and DE-FG03-96ER40969.
Section 2 was written in collaboration with Dave Soper and Francesco Hautmann.
Section 3 is based on
collaborations with I.~Schmidt and T.~Rizzo. I also thank C. J., D. Robertson,
and A. Pang for helpful conversations.

\nonumsection{References}

\end{document}